\documentclass[aps,amsfonts,preprint]{revtex4}
\def\S2{\bar{S}}
\def\a{\alpha}
\def\b{\beta}

\def\d{\delta}

\def\m{\mu}
\def\P{\Pi}

\def\o{\omega}
\def\t{\tau}
\def\an{a_{n}}
\def\ta{\tilde{a}_n}
\def\and{a_{n}^\dagger}
\def\anb{\bar{a}_n}
\def\tab{\tilde{\bar{a}}_n}
\def\on{\omega_{n}}
\def\q{S_{n}}
\def\tq{\tilde{S}_n}

\def\x{\S2_{n}}
\def\tx{\tilde{\S2}_n}
\def\sn2d{\Sn2^\dagger}
\def\({\left(}
\def\){\right)}
\def\<{\left\langle}
\def\>{\right\rangle}
\def\s{\sigma}

\begin{document}
\title{Superstring in a pp-wave background at finite temperature
\\
- TFD approach - }

\author{Daniel L. Nedel{\footnote{daniel@ift.unesp.br}}, 
M. C. B. Abdalla{\footnote {mabdalla@ift.unesp.br}}
and A. L. Gadelha{\footnote {gadelha@ift.unesp.br}}}

\affiliation{Instituto de F\'{\i}sica Te\'{o}rica, Unesp, Pamplona 145,
S\~{a}o Paulo, SP, 01405-900, Brazil }

\begin{abstract}
A thermodynamical analysis for the type IIB superstring in a pp-wave
background is considered. The thermal Fock space is built and
the temperature SUSY breaking appears naturally by analyzing the
thermal vacuum. All the thermodynamical quantities are derived by
evaluating matrix elements of operators in the thermal Fock space.
This approach seems to be suitable to study thermal effects in the
BMN correspondence context.
\end{abstract}

\maketitle

\section{Introduction}

The ADS/CFT correspondence is a concrete realization of a duality
relating gravity and Yang-Mills theory. In its strong version this
duality asserts that the ${\cal N}=4$ $SU(N)$ Super Yang Mills theory and
type IIB superstring on $ADS_{5}\times S^5$ with $N$ units of five form
flux are exactly equivalent \cite{Malda}. Although a large amount of
evidence has emerged, proving this statement turns out to be absolutely
hard to come by due to the nonlinearities of the world sheet action.
 
This scenario has changed with the discovery of a new maximally
supersymmetric solution of type IIB supergravity, the pp-waves
\cite{Blau1}. Such a background is obtained as the Penrose limit of
$ADS_{5}\times S^5$ \cite{Blau2}. On the gauge theory side the
limit  focuses on a set of operators which have R charges
$J$ and the conformal dimension $\Delta$ satisfying
$J\approx\sqrt{N}$ and $\Delta\approx J$, for fixed Yang-Mills
coupling and N going to infinity. This set up came to be known
as the BMN (Berenstein, Maldacena, Nastase) limit \cite{BMN}.
In addiction it turns out that the superstring is exactly solvable
in the pp-wave background and the above mentioned duality is
perturbatively accessible from both sides of the correspondence,
contrary to the original $ADS_{5}\times S^5$ case.

In the case of superstring at finite temperature the BMN
correspondence is suitable, since a theory which has a good
thermodynamical behaviour is related to another which has
not (string theory). Besides to the usual difficulties
in studying the thermodynamics of theories which contain gravity,
in string theory the level density of states grows exponentially, 
originating the Hagedorn temperature; at this temperature the free
energy diverges \cite{Witten2}.

Lately there have been some interesting works studying finite
temperature effects of type IIB superstring in the pp-waves
background \cite{Zayas}, \cite{Greene}, \cite{Br}, \cite{Suga},
\cite{Grig}. In a general way these works compute the superstring
partition function and the free energy on a torus. The Hagedorn
temperature is calculated using the modular
properties of the partition function. 
It was shown that the Hagedorn temperature can be
related to a phase transition \cite{Zayas}, it may be the
deconfinement/confinement phase transition on the gauge side
\cite{Br}. However, to have a complete understanding of the
superstring thermal effects in terms of gauge thermal effects,
it is necessary to understand the thermal version of
the BMN correspondence and how the temperature SUSY breaking can
affect it. Although it has been conjectured that the correspondence
exists at finite temperature \cite{Witten}, there are few efforts
to test or prove it. In this letter we present a finite
temperature formalism that can be useful for this purpose. 
  
The BMN correspondence is an equivalence of operators and Hilbert
spaces of two different theories. The central relation is
\begin{equation}
\frac{ H}{\m}\rightarrow \Delta - J \nonumber , \label{op}
\end{equation}
where  $\m$ is the only term that comes from the Ramond-Ramond
five form and survives to the BMN limit. The string hamiltonian
operator $H$ acts on the Fock space built with the string oscillators
and gives the energy of each state; $\Delta - J $ acts on the set of
gauge invariant operators which survives to BMN limit, giving their
conformal dimension minus the $J$ charge. Thus, from the operator
correspondence $\(\ref{op}\)$, we have a map between the two spaces where
the operators act upon as well as a correspondence between the matrix
elements in the basis related by this map. How can the temperature be
introduced in this scenario?

Due to the operator character of the BMN correspondence, it is useful
to introduce the temperature in such a way, that thermodynamics
quantities can be derived using operators, Hilbert spaces and matrix
elements. This is the case of Thermo Field Dynamics (TFD) developed
in \cite{UME}. TFD is a real time formalism where the main idea is to
interpret the statistical average of any quantity $Q$ over a statistical
ensemble as the expectation value of $Q$ in a thermal vacuum
\begin{equation}
 Z^{-1}(\b ){\mbox Tr}[Q e^{-\b H}]~=~
\left\langle 0 (\b ) \left| Q 
\right| 0 (\b )\right\rangle\ ,
\label{tfd}
\end{equation}
where $\b = (k_B T)^{-1}$ and $k_B$ is the Boltzmann's constant.
All the thermodynamics quantities can be defined as
matrix elements of an operator in the thermal vacuum. Also, we have
thermal operators acting upon the thermal Fock space with the same
properties of the $T=0$ ones. This comes from the fact that the
$T\neq 0$ operators and Fock space are construct from the original
$T=0$ ones by a Bogoliubov transformation. In addiction, the
temperature SUSY breaking and the respective Goldstinos come naturally
in this formalism \cite{MU,das}. 

Concerning bosonic string theory  TFD was employed to study
questions such as string field theory and renormalizability in 
\cite{YLE1}, \cite{YLE2}, \cite{FNA1}, \cite{FNA2}. The thermal
heterotic strings were presented in \cite{YLE3}, \cite{FNA3}.
Recently, in a set of constructive works \cite{IVV}, \cite{AGV2},
\cite{AEG}, \cite{AGV3}, \cite{AGV4}, \cite{AGV5}, TFD has
been used to search a microscopic description for the bosonic
$D$-branes thermodynamics.

The aim of this letter is to apply the TFD approach to construct a
thermal superstring in the pp-waves background emphasising that
the method can be useful to understand thermal effects in the BMN
correspondence.
In order to take into account the level match condition of the type IIB
superstring,  it is necessary to reformulate the expectation value of
$\(\ref{tfd}\)$ as follows:
\begin{equation}
 Z^{-1}(\b )\int_{0}^{1} d\lambda {\mbox Tr}[Q e^{-\b H+2\pi i\lambda P}]=
\int_{0}^{1} d\lambda \left\langle 0 (\b,\lambda ) \left| Q 
\right| 0 (\b,\lambda )\right\rangle,
\end{equation}
where $P$ is the momentum operator of the world sheet and the
dependence of the thermal vacuum on the lagrange multiplier comes from
the Bogoliubov transformation parameter.  
The action of type IIB superstring in a pp-wave background is only known
in the light-cone gauge. In this gauge the energy is $P^0 = P^+ +  P^-$,
where $ P^+\left |\Phi\> = p^+\left |\Phi\>$, $ P^-\left |\Phi\> =
1/p^+\(H_{lc}\)\left |\Phi\>$  and $H_{lc}$ is the light-cone hamiltonian.
As pointed out in \cite{AO}, the one string full partition function,
$Z\(\b\)= Tr e^{-\b P^0}$,
is obtained from the transverse ones $z_{lc}\(\b / p^+\)=
Tr e^{-\(\b / p^+\)H_{lc}}$,  by means of a Laplace transform:
\begin{equation}
Z\(\b\)=\frac{L}{\sqrt{2\pi}}\int dp^+ e^{\b p^+}z_{lc}\(\b/p^+\), 
\end{equation}
where $L$ is the length of the longitudinal direction. In general,
in the imaginary time formalism, the transverse partition
function is calculed by evaluating the path integral on a torus. In this
letter we will concentrated just in the transverse sector keeping $ p^+$ 
fixed and look for a thermal vacuum that reproduces the trace in this sector.
    
This work is organized as follows. The string in the pp-wave background
is described in the next section. In Section $3$ the TFD approach is
carried out to construct the thermal Fock space and the thermal operators
for this superstring. Particularly the thermal SUSY breaking and the
realization of the Goldstone theorem are showed through the analysis of
the thermal vacuum. Finally, in Section $4$, the free energy, thermal energy
and entropy of the superstring are calculated by evaluating matrix elements of
operators in the thermal Fock space; in that section the $\lambda$ dependence
of the Bogoliubov parameter appears naturally when the free energy is minimized.
In the last section conclusions and possible extensions of this work are
discussed.

\section{Type IIB Superstring on the PP Wave Background}

In this section we summarize some well known results for the type IIB
superstring on the pp-waves background. We use the light-cone
coordinates $X^\pm = \frac{1}{\sqrt{2}}\left(X^9 \pm X^0\right)$ and
we write the remaining 8 components of the spinors (after Kappa
symmetry fixing) as $S^a$, $\S2^b$, composing the ${\bf 8_{s}}$
representation of SO(8). The chiral representation of SO(8) gamma
matrices is used.

The metric of the pp-wave is
\begin{equation}
{ds}^{2}=2{dx}^{+}{dx}^{-}-{\mu}^{2}{x}^{I}{x}^{I}{dx}^{I}{dx}^{I}
+{dx}^{I}{dx}^{I},\qquad I=i, i^{\prime},
\end{equation}
$i=1,...4$, $i^{\prime}=5,...8$. It is obtained from $ADS_5 \times S^5$ by a
Penrose limit, where the only surviving components of the Ramond-Ramond five
form are:
$F_{+1234}= F_{+5678}=\mu$. This metric preserves all the 32
supersymmetries of the type IIB superstring but breaks the $SO(8)$ down
to $SO(4)\times SO(4)$. The light-cone gauge (Kappa) fixed action for
type IIB Green-Schwarz superstring on this geometry is \cite{Met1}
\begin{equation}
S=\frac{1}{2\pi\a^{\prime}}\int
{d{\sigma}}^{2}\(\frac{1}{2}\partial_{+}X^I
\partial_{-}X^I -\frac{1}{2}m^2 (X^I )^2 +iS^{a} \partial_{+}S^a +
i\S2^{a} \partial_{-}\S2^{a} -2imS^{a}\P_{ab}\S2^b\),
\end{equation}
where $\partial_{\pm} =\partial_{\tau} \pm \partial_{\sigma}$ and
$\P$ is a traceless tensor defined as $\Pi= \gamma^1 \gamma^2
\gamma^3 \gamma^4$. The mass parameter $m$ is defined as $m=\mu\
\a^{\prime} p^{+}$.
The solutions of the equations of motion with periodic boundary
conditions are
\cite{Met2}
\begin{eqnarray}
X^I = x^I_0 \cos(m\t)+\frac{\a^\prime}{m}p^I_0 \sin(m\t)\
+\sqrt{\frac{\a^\prime}{2}}
\sum_{n > 0}\frac{1}{\sqrt{\o_n}}
\left[\(\an^I e^{-i(\on\t-k_n\s)}+\an^{\dagger \: I}
e^{i(\on\t - k_n\s)}\)
\right.
\nonumber
\\
\left.
+\({\bar a}_{n}^{I} e^{-i(\on\t+k_n\s)}+{\bar a}_{n}^{\dagger \: I}
e^{i(\on\t + k_n\s)}\)\right],
\end{eqnarray}
and
\begin{eqnarray}
S^a = \cos(m\t)S_0^a + \sin(m\t)\P_{ab} \S2_0^b
+ \sum_{n > 0} c_{n} \left[\q^{a} e^{-i(\on\t-k_n\s)}\right.
\left. + S_{n}^{\dagger a} e^{i\( \o_{n}\t-k_{n}\s\)}\right.
\nonumber
\\
+ \left. i\frac{\on-k_n}{m}\P_{ab}\(\S2_{n}^{b}
e^{-i(\on\t+k_n\s)}-\S2_{n}^{\dagger b} e^{i\(\o_{n}\t+k_{n}\s\)}
\)\right],
\end{eqnarray}
\begin{eqnarray}
\S2^a = \cos(m\t)\S2_{0}^a - \sin(m\t)\P_{ab} S_{0}^b
+ \sum_{n > 0}c_{n}\left[\S2_{n}^{a} e^{-i(\on\t+k_n\s)}\right.
\left. + \S2_{n}^{\dagger a} e^{i\( \o_{n}\t+k_{n}\s\)}\right.
\nonumber
\\
- \left. i\frac{\on-k_n}{m}\P_{ab}\(S_{n}^{b}
e^{-i(\on\t-k_n\s)}- S_{n}^{\dagger b} e^{i\(\o_{n}\t-k_{n}\s\)}
 \)\right],
\end{eqnarray}
where we set:
\begin{equation}
\o_n = \sqrt{m^2 + k_n^2},\qquad c_n =\frac{1}{\sqrt{1+(\frac{\o_{n}
- k_{n}}{m})^2}},\qquad k_n =2\pi n.
\end{equation}

The canonical quantization gives the standard commutator and
anti-commutator relations of harmonic oscillator for $\an$,
$\an^{\dagger}$ and $\q$, $\q^\dagger$, respectively, and the same
for ``bar'' operators. The zero mode part is written as follows
\begin{eqnarray}
a_0^I &=& \frac{1}{\sqrt{2m}}(p_0^I -imx_0^I),\qquad
a_0^{\dagger\:I}=
\frac{1}{\sqrt{2m}}(p_0^I +imx_0^I),\nonumber \\
S_{\pm}^a &=&\frac{1}{2}\(1\pm \P\)_{ab}\frac{1}{\sqrt{2m}}\(S_0^b
\pm i\S2_0^b\),\qquad S_{\pm}^{\dagger\:a}= \frac{1}{2}\(1\pm
\P\)_{ab}\frac{1}{\sqrt{2m}}\(S_0^b \mp i\S2_0^b\),
\end{eqnarray}
which satisfies
\begin{eqnarray}
\left[a_0^I,a_0^{\dagger\:J}\right]&=& \d^{IJ}, \qquad
\left[a_0^{\dagger\:I},a_0^{\dagger\:J}\right]=
\left[a_0^I,a_0^J\right]=0, \nonumber \\
\left\{ S_{\pm}^a, S_{\pm}^{\dagger\:b}\right\}&=&\d^{ab},\qquad
\left\{ S_{\pm}^a, S_{\pm}^b\right\}=\left\{ S_{\pm}^{\dagger\:a},
S_{\pm}^{\dagger\:b}\right\}=0.
\end{eqnarray}

The light-cone hamiltonian is calculated in a standard way and it
is written as
\begin{equation}
p_{+} H= m\(a_0^{\dagger\:I}a_{0}^{I} + S_{+}^{\dagger\:a}S_{+}^a +
S_{-}^{\dagger\:a}S_{-}^a \)+ \sum_{n>0}\on\(\an^{\dagger\:I}\an^{I}
+\anb^{\dagger\:I}\anb + \q^{\dagger\:a}\q^{a} +
\x^{\dagger\:a}\x^a\).
\end{equation}

In addiction to the time translations generated by the hamiltonian, the
action has $29$ more bosonic symmetries (generated by $P^+$, $P^I$ and by
the rotations $J^{+ I}$, $J^{ij}$, $J^{i^{\prime}, j^{\prime}}$) and $32$
supersymmetries.
The fermionic set of generators has $16$ kinematical supercharges,
that belong to ${\bf 8_{s}}$ of $SO(8)$ and changes the
polarizations of the fields. The remaining
$16$ fermionic symmetries are the dynamical supercharges, that
transform the fields of the same supermultiplet and belong to
${\bf 8_{c}}$ of $SO(8)$. While the kinematical supercharge does
not commute with the hamiltonian, the dynamical one does, providing
a supersymmetric spectrum. The dynamical supercharges can be
written as $Q_{\dot\a}^\pm = Q_{\dot\a}\pm \bar{Q}_{\dot\a}$,
where
\begin{eqnarray}
\frac{\sqrt{p^+}}{2^{1/4}}Q_{\dot a}&=&
P_{0}^I\(\gamma^I S_{0}\)_{\dot a} -
mX_0^I\(\gamma \Pi \bar{S}_0\)_{\dot a}
\nonumber 
\\
&+& \sum_{n>0}\left[( \sqrt{2\on}c_n\(\an^{\dagger\:I}\gamma^I\q +
\an^I\gamma^I\q^{\dagger}\)_{\dot a} +
\frac{im}{\sqrt{2\on}c_n}\(\gamma\P\)_{{\dot a}b}(\anb^{\dagger\:I}\x^b
- \anb^I\x^{\dagger\:b})\right], \nonumber \\
\end{eqnarray}
and $\bar{Q}_{\dot{a}}$  can be obtained from $Q_{\dot{a}}$ replacing
``bar'' variables by non-bar variables and $i$ by $-i$,  while the
kinematic supercharges are $Q\approx S_0 $ and
$\bar{Q}\approx \bar{S}_0 $. 
  
Finally, we can choose the vacuum $\left|0,p^{+}\>$ as defined by
\begin{eqnarray}
\q\left|0,p^+\>&=&\x\left|0,p^+\>=0 \qquad n>0,
\nonumber \\
a_n^I\left|0,p^+\>&=&{\bar a}_n^I\left|0,p^+\>=0 \qquad n>0,
\nonumber \\
S_{\pm}\left|0,p^+\>&=&a_{0}^{I}\left|0,p^+\>=0.
\end{eqnarray}
The hamiltonian and the dynamical supercharges annihilate the vacuum
as a signal of supersymmetry. Following the BMN dictionary, the
vacuum has zero energy and is related to an operator in the gauge side
with zero value for $\Delta - J$ :
\begin{equation}
\left|0,p^+\> \rightarrow O^J(0) \left|vac\>, \qquad
O(x)=\frac{1}{\sqrt{JN^J}}TrZ^J,
\end{equation}
where $\left|vac\>$ is the Yang-Mills vacuum and $O^J$ is composed of
two out of the six scalar fields of the $N=4$ super Yang Mills multiplet:
$ Z= \frac{1}{2}\(\phi^5 + i\phi^6\)$. The trace is taken over the
$SU(N)$ index. The next section is devoted to construct a thermal vacuum
for the string, that can be useful to understand how the above dictionary
is affected by the temperature.

\section{TFD Approach}

Let us now apply the TFD approach to construct the thermal Fock space
for superstring on a pp-wave background. Following Umezawa, to
provide enough room to accommodate the thermal properties of the
system, we have first to duplicate the degrees of freedom. To this
end we construct a copy of the original Hilbert space,
denoted by $\widetilde{H}$. The Tilde Hilbert space is built with a set of
oscillators: $\tilde{a}_0$, $\tilde{S}_{\pm}$, $\ta$, $\tab$, $\tq$,
$\tx$ that have the same (anti-) commutation properties as the original
ones. The operators of the two systems commute among themselves and the
total Hilbert space is the tensor product of the two spaces.

We can now construct the thermal vacuum. This is achieved by implementing
a Bogoliubov transformation in the total Hilbert space. The transformation
generator is given by
\begin{equation}
G=G^{B}+G^{F}, \label{gen}
\end{equation}
for
\begin{eqnarray}
G^{B}&=&G_{0}^{B} + \sum_{n=1} \(G_{n}^{B} + {\bar G}_n^{B}\),
\label{genb}
\\
G^{F}&=&G_{+}^{F} + G_{-}^{F} + \sum_{n=1} \(G_{n}^{F} + {\bar
G}_{n}^{F}\),
\label{genf}
\end{eqnarray}
where
\begin{eqnarray}
G_{0}^{B}&=&-i\theta_{0}^{B}\(a_{0}\cdot {\tilde a}_{0}-{\tilde
a}_{0}^{\dagger} \cdot a_{0}^{\dagger}\),
\label{gen0b}
\\
G_{n}^{B}&=&-i\theta_{n}^{B}\(a_{n}\cdot {\tilde a}_{n}-{\tilde
a}_{n}^{\dagger} \cdot a_{n}^{\dagger}\),
\label{genbn}
\\
{\bar G}_{n}^{B}&=&-i{\bar \theta}_{n}^{B}\({\bar a}_{n}\cdot {\tilde
{\bar a}}_{n}-{\tilde {\bar a}}_{n}^{\dagger} \cdot {\bar
a}_{n}^{\dagger}\),
\label{genbnb}
\\
G_{\pm}^{F}&=&-i\theta_{\pm}^{F}\({\widetilde S}_{\pm}\cdot
S_{\pm}-S_{\pm}^{\dagger} \cdot {\widetilde S}_{\pm}^{\dagger}\),
\label{gen0f}
\\
G_{n}^{F}&=&-i\theta_{n}^{F}\({\widetilde S}_{n}\cdot
S_{n}-S_{n}^{\dagger} \cdot {\widetilde S}_{n}^{\dagger}\),
\label{genfn}
\\
G_{n}^{F}&=&-i{\bar \theta}_{n}^{F}\({\widetilde {\bar S}}_{n}\cdot
{\bar S}_{n}-{\bar S}_{n}^{\dagger} \cdot {\widetilde {\bar
S}}_{n}^{\dagger}\).
\label{genfnb}
\end{eqnarray}
Here, the labels $B$ and $F$ specifies fermions and bosons,
the dots represent the inner products and $\theta$, ${\bar \theta}$
are real parameters. In the thermal equilibrium they are related to
the Bose-Einstein and Fermi-Dirac distribution of the oscillator
$n$ as we will see. The thermal vacuum is given by the following relation
\begin{eqnarray}
\left |0\(\theta\)\right\rangle &=& e^{-i{G}}\left.
\left|0\right\rangle \!\right\rangle \nonumber
\\
&=& \left( \frac{1}{\cosh(\theta_{0}^{B})}\right)^{8}
\left(\cos(\theta_{+}^{F})\right)^{4}\left(\cos(\theta_{-}^{F})\right)^{4}
e^{\tanh\(\theta_{0}^{B}\)\(a_{0}^{\dagger}\cdot {\tilde
a}_{0}^{\dagger}\)}e^{\tan\(\theta_{+}^{F}\)\(S_{+}^{\dagger} \cdot
{\widetilde S}_{+}^{\dagger}\) +
\tan\(\theta_{+}^{F}\)\(S_{-}^{\dagger} \cdot {\widetilde
S}_{-}^{\dagger}\)} \nonumber
\\
& \times &\prod_{n=1}\left[\left(
\frac{1}{\cosh(\theta_{n}^{B})}\right)^{8}\left( \frac{1}{\cosh({\bar
\theta}_{n}^{B})}\right)^{8}
e^{\tanh\(\theta_{n}^{B}\)\(a_{n}^{\dagger}\cdot {\tilde
a}_{n}^{\dagger}\)+ \tanh\({\bar \theta}_{n}^{B}\)\({\bar
a}_{n}^{\dagger}\cdot {\tilde {\bar a}}_{n}^{\dagger}\)}\right.
\nonumber
\\
&\times& \left.\left(\cos(\theta_{n}^{F})\right)^{8}\left(\cos({\bar
\theta}_{n}^{F})\right)^{8}e^{\tan\(\theta_{n}^{F}\)\(S_{n}^{\dagger}
\cdot {\widetilde S}_{n}^{\dagger}\) + \tan \({\bar
\theta}_{n}^{F}\)\({\bar S}_{n}^{\dagger} \cdot {\widetilde {\bar
S}}_{n}^{\dagger}\)}\right] \left.
\left|0\right\rangle\!\right\rangle \label{tva}.
\end{eqnarray}

The creation and annihilation operators at $T \neq 0$ are given by
the Bogoliubov transformation as follows
\begin{eqnarray}
S_{\pm}^{a}\(\theta_{\pm}^{F}\)&=&e^{-iG}S_{\pm}^{a}e^{iG}
=\cos\(\theta_{\pm}^{F}\)S_{\pm}^{a} -
\sin\(\theta_{\pm}^{F}\){\widetilde S}_{\pm}^{\dagger \: a},
\\
S_{n}^{a}\(\theta_{n}^{F}\)&=&e^{-iG}S_{n}^{a}e^{iG}
=\cos\(\theta_{n}^{F}\)S_{n}^{a} - \sin\(\theta_{n}^{F}\){\widetilde
S}_{n}^{\dagger \: a},
\\
{\bar S}_{n}^{a}\({\bar \theta}_{n}^{F}\)&=&e^{-iG}{\bar
S}_{n}^{a}e^{iG} =\cos\({\bar \theta}_{n}^{F}\){\bar S}_{n}^{a} -
\sin\({\bar \theta}_{n}^{F}\){\widetilde {\bar S}}_{n}^{\dagger \: a},
\\
a_{0}^{I}\(\theta_{0}^{B}\)&=&e^{-iG}a_{0}^{I}e^{iG}
=\cosh\(\theta_{0}^{B}\)a_{0}^{I} -
\sinh\(\theta_{0}^{B}\){\widetilde a}_{0}^{\dagger \: I},
\\
a_{n}^{I}\(\theta_{n}^{B}\)&=&e^{-iG}a_{n}^{I}e^{iG}
=\cosh\(\theta_{n}^{B}\)a_{n}^{I} -
\sinh\(\theta_{n}^{B}\){\widetilde a}_{n}^{\dagger \: I},
\\
{\bar a}_{n}^{I}\({\bar \theta}_{n}^{B}\)&=&e^{-iG}{\bar
a}_{n}^{I}e^{iG} =\cosh\({\bar \theta}_{n}^{B}\){\bar a}_{n}^{I} -
\sinh\(\theta_{n}^{B}\){\widetilde {\bar a}}_{n}^{\dagger \: I}.
\end{eqnarray}
These operators annihilate the state written in $\(\ref{tva}\)$ defining
it as the vacuum. The creation operators are obtained from the above
list by hermitian conjugation. The tilde counterparts
can be obtained using the tilde conjugation rules defined in
\cite{UME}. As the transformation generator defined in
$\(\ref{gen}\)-\(\ref{genfnb}\)$ is hermitian and changes the signal
under tilde conjugation, one can check that the vacuum is invariant
under this conjugation.

The thermal Fock space is constructed from the vacuum
$\(\ref{tva}\)$ by applying the thermal creation operators.
As the Bogoliubov transformation is canonical, the thermal
operators obey the same (anti-) commutators relations as the
operators at $T=0$.

The hamiltonian plays an important r\^{o}le in the BMN correspondence,
so a natural question is what is the r\^{o}le it plays in the thermal
Fock space. It is easy to see that thermal states are not
eigenstates of the original hamiltonian but they are eigenstates
of the combination:
\begin{equation}
{\widehat H} = H - {\widetilde H},
\label{hath}
\end{equation}
in such a way that ${\widehat H}$ plays the
r\^{o}le of the hamiltonian generating the temporal translation in
the thermal Fock space. Using the commutation relations we can
prove that the Heisenberg equations are satisfied replacing $H$ and
${\widetilde H}$ by ${\widehat H}$. Also we have
${\widehat Q}_{\dot{a}}$ and ${\widehat {\bar Q}}_{\dot{a}}$
defined in a similar way as ${\widehat H}$ in $\(\ref{hath}\)$.
These new supercharges realize the same supersymmetry algebra as
the supercharges at $T=0$. However, this fact does not imply that
the supersymmetry remains at finite temperature. In fact,
in the TFD approach, thermal effects are observed when one considers
$T=0$ operators expectation values on the thermal Fock space.
For $T=0$ the dynamical supercharge commutes with the hamiltonian
and as a consequence of the supersymmetry algebra annihilates
the vacuum. Now, by applying the $Q_{\dot{a}}^+$ operator on the
thermal vacuum, we get:
\begin{eqnarray}
\frac{ Q_{\dot{a}}^+}{2^{1/4}\sqrt{\m}}
\left |0\(\theta\)\right\rangle &=&
\left[ a_0^{\dagger\:I}\(\theta_{0}^{B}\)
\(\gamma^I\tilde{S}_+^{\dagger}\(\theta_{+}^{F}\)\)_{\dot a}
\cosh\(\theta_0^B\)\sin\(\theta_{+}^{F}\)
\right.
\nonumber
\\
&&\left.
+ {\tilde{a}}_0^I\(\theta\)
\(\gamma^I S_{-}^{\dagger}\(\theta_{-}^{F} \) \)_{\dot{a}}
\sinh\(\theta_0^B\)\cos\(\theta_{-}^{F}\)
+{\rm osc. terms} \right] \left|0\(\theta\)\right\rangle .
\end{eqnarray}
These new excitations generated by $Q_{\dot a}$ have interesting
properties. If we apply the hamiltonian ${\widehat H}$ in this state we
have:
\begin{equation}
{\widehat H}\(Q_{\dot{a}}\left |0\(\theta\)\right\rangle\)=0,
\end{equation}
so these excitations are in fact massless excitations with respect to
the hamiltonian ${\widehat H}$. For each supersymmetric
oscillator $n$ we have a massless excitation proportional to 
$ a_n^{\dagger\:I}\(\theta_{n}^{B}\)
\(\gamma^I \tilde{S}_n\(\theta_{n}^{F}\)\)+
\tilde{a}_n\(\theta_n^{B}\)^{\dagger\:I}
\(\gamma^I S_n^{\dagger}\(\theta_{n}^{F}\)\)$
plus ``bar'' variables. This combination is called super pair and
realizes the Goldstone theorem for the supersymmetry breaking
generated by the temperature \cite{MU,das}. They play the r\^{o}le of the
Goldstinos, although they are not really particles since there are no 
interactions.

In this section we have constructed both the thermal vacuum and
Fock space and demonstrated the breaking of $T=0$ SUSY. Next
section is devoted to find thermodynamic quantities and analyze the
thermodynamics of the superstring on a pp-wave background.

\section{Thermodynamical analysis}

In this section the TFD approach will be used to compute
thermodynamical quantities by evaluating matrix elements of operators
in the thermal Fock space. It was appointed out by Polchinski \cite{Pol}
that, in the one-string sector, the torus path integral computation
of the free energy coincides with what we would obtain by adding the
contributions from different states of the spectrum to the free energy.
Here the free energy is obtained from the knowledge of the
thermal energy and entropy operators.    

The energy operator is such that the level matching condition
must be implemented. The way we proceed is to consider
a shifted hamiltonian in the sense of Ref. \cite{OJI}, as follows
\begin{equation}
H=\frac{1}{p+}\left[mN_{0}+\sum_{n=1}\omega _{n}\left( N_{n}+\bar{N}%
_{n}\right)\right] +\frac{1}{\beta}i \lambda
\sum_{n=1}k_{n}\left( N_{n}-\bar{N}_{n}\right),
\label{hshif}
\end{equation}
where
\begin{equation}
N_{0}=N_{0}^{B}+N_{+}^{F}+N_{-}^{F},
\end{equation}
and
\begin{equation}
N_{n}=N_{n}^{B}+N_{n}^{F},\qquad \bar{N}_{n}=\bar{N}_{n}^{B}+
\bar{N}_{n}^{F}.
\end{equation}
Computing the expectation value of $\(\ref{hshif}\)$ in the thermal vacuum
$\(\ref{tva}\)$, the following result arises
\begin{eqnarray}
E && \equiv \int_{0}^{1}d\lambda \left\langle 0\left( \theta_{\lambda} \right) 
\right| H\left| 0\left( \theta_{\lambda} \right) \right\rangle  \nonumber
\\
&&=\int_{0}^{1}d\lambda \left[ 
\frac{m}{p+}\left[ 8\sinh^{2}\left( \theta _{0}^{B}\right) +4\sin^{2}\left(
\theta _{+}^{F}\right) +4\sin^{2}\left( \theta _{-}^{F}\right)\right] 
\right.
\nonumber
\\
&&\left. +\frac{8}{p+}\sum_{n=1}\omega _{n}\left[ \sinh^{2}\left( \theta
_{n}^{B}\right) +\sin^{2}\left( \theta _{n}^{F}\right) +\sinh^{2}\left(
\overline{\theta }_{n}^{B}\right) +\sin^{2}\left( \overline{\theta }%
_{n}^{F}\right) \right]\right.
\nonumber
\\
&&\left. + \frac{8 \lambda i}{\beta} \left[ \sum_{n=1}k_{n}\left(
\sinh^{2}\left( \theta _{n}^{B}\right) +\sin^{2}\left( \theta _{n}^{F}\right)
-\sinh^{2}\left( \overline{\theta }_{n}^{B}\right) -\sin^{2}\left( \overline{%
\theta }_{n}^{F}\right) \right) \right] \right],
\end{eqnarray}
where $\theta_{\lambda}$ just specifies the lagrange multiplier dependence
of the Bogoliubov parameter. Concerning the entropy operator, an extension to
that presented in \cite{AGV2} is necessary in order to include the fermionic
degrees of freedom. Namely,
\begin{equation}
K=K^{B}+K^{F},
\end{equation}
where the boson contribution is given by
\begin{eqnarray}
K^{B} &=&-\left\{ a_{0}^{\dagger }\cdot a_{0}\ln \left( \sinh^{2}\left(
\theta _{0}^{B}\right) \right) -a_{0}\cdot a_{0}^{\dagger }\ln \left( \cosh
^{2}\left( \theta _{0}^{B}\right) \right) \right\}
\nonumber
\\
&&-\sum_{n=1}\left\{ a_{n}^{\dagger }\cdot a_{n}\ln \left( \sinh^{2}\left(
\theta _{n}^{B}\right) \right) -a_{n}\cdot a_{n}^{\dagger }\ln \left( \cosh
^{2}\left( \theta _{n}^{B}\right) \right) \right\}
\nonumber
\\
&&-\sum_{n=1}\left\{ \overline{a}_{n}^{\dagger }\cdot \overline{a}_{n}\ln
\left( \sinh^{2}\left( \overline{\theta }_{n}^{B}\right) \right) -\overline{a}%
_{n}\cdot \overline{a}_{n}^{\dagger }\ln \left( \cosh ^{2}\left( \overline{%
\theta }_{n}^{B}\right) \right) \right\},
\end{eqnarray}
and
\begin{eqnarray}
K^{F} &=&-\left\{ S_{+}^{\dagger }\cdot S_{+}\ln \left( \sin^{2}\left( \theta
_{+}^{F}\right) \right) +S_{+}\cdot S_{+}^{\dagger }\ln \left( \cos
^{2}\left( \theta _{+}^{F}\right) \right) \right\}
\nonumber
\\
&&-\left\{ S_{-}^{\dagger }\cdot S_{-}\ln \left( \sin^{2}\left( \theta
_{-}^{F}\right) \right) +S_{-}\cdot S_{-}^{\dagger }\ln \left( \cos
^{2}\left( \theta _{-}^{F}\right) \right) \right\}
\nonumber
\\
&&-\sum_{n=1}\left\{ S_{n}^{\dagger }\cdot S_{n}\ln \left( \sin^{2}\left(
\theta _{n}^{F}\right) \right) +S_{n}\cdot S_{n}^{\dagger }\ln \left( \cos
^{2}\left( \theta _{n}^{F}\right) \right) \right\}
\nonumber
\\
&&-\sum_{n=1}\left\{ \overline{S}_{n}^{\dagger }\cdot \overline{S}_{n}\ln
\left( \sin^{2}\left( \overline{\theta }_{n}^{F}\right) \right) +\overline{S}%
_{n}\cdot \overline{S}_{n}^{\dagger }\ln \left( \cos ^{2}\left( \overline{%
\theta }_{n}^{F}\right) \right) \right\},
\end{eqnarray}
is the entropy operator for the fermionic sector.
The evaluation of the entropy operator on the thermal vacuum leads to 
the following result
\begin{eqnarray}
S &\equiv&\int_{0}^{1}d\lambda \left\langle 0\left( \theta_{\lambda} \right)
\right| K \left| 0\left( \theta_{\lambda} \right) \right\rangle  
\nonumber
\\
&=&\int_{0}^{1}d\lambda \left\{ -8\left[ \sinh^{2}\left( \theta
_{0}^{B}\right) \ln \left( \tanh^{2}\left( \theta _{0}^{B}\right) \right)
-\ln \left( \cosh ^{2}\left( \theta _{0}^{B}\right) \right) \right] \right. 
\nonumber
\\
&&\left. -4\left[ \sin^{2}\left( \theta _{+}^{F}\right) \ln \left(
\tan^{2}\left( \theta _{+}^{F}\right) \right) +\ln \left( \cos ^{2}\left(
\theta _{+}^{F}\right) \right) \right] \right.  
\nonumber
\\
&&\left. -4\left[ \sin^{2}\left( \theta _{-}^{F}\right) \ln \left(
\tan^{2}\left( \theta _{-}^{F}\right) \right) +\ln \left( \cos ^{2}\left(
\theta _{-}^{F}\right) \right) \right] \right.  
\nonumber
\\
&&\left. -8\sum_{n=1}\left[ \sinh^{2}\left( \theta _{n}^{B}\right) \ln \left(
\tanh^{2}\left( \theta _{n}^{B}\right) \right) -\ln \left( \cosh ^{2}\left(
\theta _{n}^{B}\right) \right) \right] \right.  
\nonumber
\\
&&\left. -8\sum_{n=1}\left[ \sinh^{2}\left( \overline{\theta }_{n}^{B}\right) \ln
\left( \tanh^{2}\left( \overline{\theta }_{n}^{B}\right) \right) -\ln \left(
\cosh ^{2}\left( \overline{\theta }_{n}^{B}\right) \right) \right] \right. 
\nonumber
\\
&&\left. -8\sum_{n=1}\left[ \sin^{2}\left( \theta _{n}^{F}\right) \ln \left(
\tan^{2}\left( \theta _{n}^{F}\right) \right) +\ln \left( \cos ^{2}\left(
\theta _{n}^{F}\right) \right) \right] \right.  
\nonumber
\\
&&\left. -8\sum_{n=1}\left[ \sin^{2}\left( \overline{\theta }_{n}^{F}\right)
\ln \left( \tan^{2}\left( \overline{\theta}_{n}^{F}\right) \right) +\ln
\left( \cos ^{2}\left( \overline{\theta }_{n}^{F}\right) \right) \right]
\right\}. 
\end{eqnarray}

Now one can construct the potential
\begin{equation}
F=E-\frac{1}{\beta }S,
\end{equation}
where explicitly we have the following expression
\begin{eqnarray}
F &=&\int_{0}^{1}d\lambda \left\{ \frac{m}{p+}\left[ 8\sinh^{2}\left( \theta
_{0}^{B}\right) +4\sin^{2}\left( \theta _{+}^{F}\right) +4\sin^{2}\left(
\theta _{-}^{F}\right) \right] \right.  
\nonumber
\\
&&\left. +\frac{1}{\beta }\left[ 8\sinh^{2}\left( \theta _{0}^{B}\right) \ln
\left( \tanh^{2}\left( \theta _{0}^{B}\right) \right) -8\ln \left( \cosh
^{2}\left( \theta _{0}^{B}\right) \right) +4\sin^{2}\left( \theta
_{+}^{F}\right) \ln \left( \tan^{2}\left( \theta _{+}^{F}\right) \right)
 \right. \right. 
\nonumber
\\
&&\left. \left. +4\ln \left( \cos^{2}\left( \theta _{+}^{F}\right) \right)
+4\sin^{2}\left( \theta _{-}^{F}\right) \ln \left(
\tan^{2}\left( \theta _{-}^{F}\right) \right) +4\ln \left( \cos ^{2}\left(
\theta _{-}^{F}\right) \right) \right] \right.  
\nonumber
\\
&&\left. +8\sum_{n=1}\left\{ \left( \frac{\omega _{n}}{p+}+\frac{i \lambda
k_{n}}{\beta }\right) \left[ \sinh^{2}\left( \theta _{n}^{B}\right)
+\sin^{2}\left( \theta _{n}^{F}\right) \right] 
\right. \right.  
\nonumber
\\
&&\left. \left. 
+\left( \frac{\omega _{n}}{p+}-
\frac{i \lambda k_{n}}{\beta }\right) 
\left[ \sinh^{2}\left( \overline{\theta }
_{n}^{B}\right) +\sin^{2}\left( \overline{\theta }_{n}^{F}\right) \right]
+\frac{1}{\beta }\left[ \sinh^{2}\left( \theta
_{n}^{B}\right) \ln \left( \tanh^{2}\left( \theta _{n}^{B}\right) \right)
 \right. \right. \right.  
\nonumber
\\
&&\left. \left. \left. 
-\ln \left( \cosh ^{2}\left( \theta _{n}^{B}\right) \right) +\sinh^{2}\left( 
\overline{\theta }_{n}^{B}\right) \ln \left( \tanh^{2}\left( \overline{\theta 
}_{n}^{B}\right) \right)
\right. \right. \right. 
\nonumber
\\
&&\left. \left. \left.
-\ln \left( \cosh ^{2}\left( \overline{\theta }
_{n}^{B}\right) \right) +\sin^{2}\left( \theta _{n}^{F}\right) \ln \left(
\tan\left( \theta _{n}^{F}\right) \right) 
\right. \right. \right.
\nonumber
\\
&&\left. \left. \left.
+\ln \left( \cos ^{2}\left( \theta
_{n}^{F}\right) \right) +\sin^{2}\left( \overline{\theta }_{n}^{F}\right) \ln
\left( \tan\left( \overline{\theta }_{n}^{F}\right) \right) +\ln \left( \cos
^{2}\left( \overline{\theta }_{n}^{F}\right) \right) \right] \right\}
\right\}.
\label{fexp}
\end{eqnarray}

Minimizing the potential $F$ with respect to $\theta$
we find the explicit dependence of these parameters in relation to
$\omega_{n}$, $\beta$ and $\lambda$. In this way we have 
\begin{equation}
\sinh^{2}\(\theta_{0}^{B}\)=\frac{1}{e^{\frac{\beta m}{p+}}-1},
\qquad
\sin^{2}\({\theta}_{\pm}^{F}\)=\frac{1}{e^{\frac{\beta m}{p+}}-1},
\end{equation}
for the zero modes, and
\begin{eqnarray}
\sinh^{2}\(\theta_{n}^{B}\)&=&\frac{1}{e^{\frac{\beta \omega_{n}}{p+}+
i\lambda k_{n}}-1},
\qquad
\sinh^{2}\( \overline{\theta}_{n}^{B}\)=\frac{1}{e^{\frac{\beta\omega_{n}}{p+}-
i\lambda k_{n}}-1},
\nonumber
\\
\sin^{2}\(\theta_{n}^{F}\)&=&\frac{1}{e^{\frac{\beta \omega_{n}}{p+}+
i\lambda k_{n}}+1},
\qquad
\sin^{2}\( \overline{\theta}_{n}^{F}\)=\frac{1}{e^{\frac{\beta\omega_{n}}{p+}-
i\lambda k_{n}}+1},
\end{eqnarray}
for the others. These expressions fix the thermal vacuum $\(\ref{tva}\)$
as those that reproduce the trace over the transverse sector.

Note that substituting the above results in expression $\(\ref{fexp}\)$ we find
\begin{equation}
F=-\frac{1}{\beta}\int_{0}^{1}d\lambda \ln \prod_{n={\mathbb Z}}
\left[ \frac{1+e^{-\frac {\beta \omega_{n}}{p+}+ i \lambda k_{n}}}
{1-e^{-\frac {\beta \omega_{n}}{p+}+i \lambda k_{n}}}\right]^{8}. \label{final}
\end{equation}
This expression is the TFD answer for the transverse free energy. To make contact
with other formalisms, define a potential $f\(\lambda\)$ such that 
\begin{equation}
F=\int_{0}^{1}d\lambda f\(\lambda\),
\end{equation}
where $f\(\lambda\)$, given by the equation $\(\ref{final}\)$, can be written
as
\begin{equation}
f\(\lambda\)=-\frac{1}{\b} \ln \(z_{lc}(\b / p^+,\lambda)\),
\end{equation}
and $z_{lc}(\b / p^+,\lambda)$ is the tranverse partition function calculated
in \cite{Zayas}.

\section{Conclusions}

In this letter we show that the Thermo Field Dynamics approach is 
very useful to calculate thermodynamical quantities for the
superstring in a pp-wave background. The main characteristic of this
approach is the construction of a thermal Fock space and thermal
operators. Owing to the operator characteristics of BMN correspondence,
the TFD approach can be a powerful method to set up a possible thermal BMN
correspondence. 

The transverse thermal energy of the superstring is derived by
evaluating the matrix elements of the $T= 0$ hamiltonian in the
thermal vacuum. In the same way the entropy and free energy are
computed. These results can be obtained in the imaginary time
formalism by evaluating the partition function on the torus. 
Here it is necessary to emphasize that in the TFD approach
the world sheet in defined on the sphere, and our results
can be compared with those coming from the torus as a consequence
of the thermal Bogoliubov transformation. This seems to avoid the problems
of the Wick rotation on a pp-wave background pointed out, for example,
in \cite{Russo}.

As a consequence of the thermal breaking of supersymmetry, the dynamical
supercharge does not annihilate the vacuum anymore. We have shown that the
dynamical supercharge excites massless modes in the vacuum, that play the
r\^{o}le of the Goldstinos \cite{MU,das}.

There are many possible extensions of this work. The most direct one
is to use the TFD algorithm to construct a thermal Hilbert space on
the gauge side of the BMN correspondence. We expect to be able to
construct a thermal Fock space and a well defined
${\hat \Delta} - {\hat J}$ operator that can be related with the
${\widehat H}$ presented in $\(\ref{hath}\)$. On the string side we can
use the TFD perturbation theory
defined in \cite{UME} to go beyond the one-string sector and include
the string field cubic interactions \cite{Spr1}, \cite{Spr2}, \cite{ari1},
\cite{ari2}, which
is a very hard task in the usual imaginary time formalism.
In another direction, we have shown that by
evaluating expected values on a sphere, we reproduce the usual torus
results of the imaginary time formalism.
These results correspond to the sum of the free energy of
each string mode, as was pointed out in \cite{Pol}.
We can use further the operator method defined in \cite{Gaume} to
evaluate the expectation values on a torus and get quantum string
corrections to the thermal energy derived in this letter. Finally, as this
formalism is a real time formalism, it is possible go out of thermal
equilibrium, which can be very useful to understand the thermodynamics
of the early universe.

\section*{Acknowledgements}
We would like to thank D. Z. Marchioro, B. M. Pimentel 
and I. V. Vancea for usefull discussions and especially A. E. Santana for
attention and comments about the work.
M. C. B. A. was partially supported by the CNPq Grant 302019/2003-0,
A. L. G. and D. L. N. are supported by a FAPESP post-doc fellowship.

\end{document}